\documentstyle[sprocl]{article}

\begin{document}

\title{VARIATIONAL APPROACH TO HYDRODYNAMICS \\ 
- FROM QGP TO GENERAL RELATIVITY}

\author{H.-TH. ELZE, T. KODAMA}

\address{Instituto de F\'{\i}sica, UFRJ, CP 68528, 
21945-970 Rio de Janeiro, Brasil} 

\author{Y. HAMA}

\address{Instituto de F\'{i}sica, Universidade de S\~{a}o Paulo, 
CP 66318, \\
05389-970 S\~{a}o Paulo, Brasil}

\author{M. MAKLER}

\address{Centro Brasileiro de Pesquisas F\'{i}sicas, 
Rua Xavier Sigaud 150, \\
22290-160 Rio de Janeiro, RJ, Brasil}

\author{J. RAFELSKI}

\address{Physics Department, University of Arizona, Tucson, AZ\,85721,
U.S.A.}

\maketitle
\abstracts{
We derive the special and general relativistic hydrodynamic 
equations of motion for ideal fluids from a variational principle. 
Our approach allows to find approximate solutions, 
whenever physically motivated trial functions can be used.  
Illustrating this, a Rayleigh-Plesset type equation for the 
relativistic motion of a spherical (QGP) droplet is obtained. Also the 
corresponding general relativistic effective Lagrangian for spherically 
symmetric systems is presented.}
\begin{center}
{\it Dedicated to the memory of Peter A. Carruthers}
\end{center}

\section{Introduction}
First applications of relativistic hydrodynamics to the process of
multiparticle production in high-energy hadronic collisions were 
studied by Fermi and Landau in the early 1950's.\,\cite{Fermi,Landau}
Recently, the relativistic motion of fluids has been studied extensively
with respect to the analysis of relativistic heavy-ion collisions.\,\cite{review} The hydrodynamic description of high-energy
hadronic and nuclear collisions has been successful in reproducing many  
characteristic features of these processes, such as multiplicity and transverse energy distributions. 

Theoretically, however, the foundation
of the hydrodynamical picture for these processes is not fully understood.
The application of hydrodynamics implicitly assumes
local thermal equilibrium via an equation of state. Therefore,
the relaxation time and the mean free path
should be much smaller compared to, respectively, the hydrodynamical time
scale and spatial size of the system, which may be satisfied best in 
collisions of heavy nuclei.

From a kinematical point of view, besides the equation of state, the equations of hydrodynamics simply express the 
conservation laws of energy, momentum, and possibly 
various charges. 
Thus, for processes where flow is an important factor, a hydrodynamic  
description seems natural to begin with. Effects of finite relaxation 
time and mean-free path
may be implemented later on, viscosity and heat 
conductivity in particular, or some simplified transport equations
might be applicable, for example, see Ref.\,\cite{EH89}.

Relativistic hydrodynamics is a local description of the
conservation laws, i.e. in terms of the energy-momentum tensor,  
\begin{equation} \label{T1}
\partial _{\mu}T^{\mu\nu}=0
\;\;, \end{equation}
where we momentarily assume cartesian coordinates. It is difficult 
to solve this set of coupled partial differential equations in 
any generality. Few analytical solutions are known, but even
for simple geometries, like one-dimensional or spherically symmetric
cases, one often has to resort to numerical solutions. 
Under these circumstances a physically motivated and more qualitative approach, such as via a variational formulation, should provide useful 
insight, before or without full scale computer simulations.  

Another important arena for relativistic hydrodynamics is
found in cosmology and high-energy astrophysics, such as the gravitational
collapse of stellar cores forming a neutron star or a black hole,
relativistic blast waves in models of gamma ray bursts, etc.\,\cite
{Cosmos,Astro} The assumption of local thermodynamical
equilibrium is considered to be well justified in most cases, 
cf. Ref.\,\cite{Retard}. However, often 
the fluid dynamics and the local changes of
the gravitational field have to be considered simultaneously. Again, this  
is generally very complicated and, therefore, a variational scheme seems  extremely helpful. 

Frequently even the equation of state of
matter is not known precisely. Rather, hydrodynamics is 
employed to infer its properties involved in the process. In
such cases we need to describe the flow, which 
characterizes the dynamics, assuming a specific equation of state.

In conclusion, a variational method is highly desirable which allows to 
solve the dynamical equations of the system more effectively. 
Such an effective Lagrangian and variational principle has previously 
been employed to study the effect of local turbulent motion on the 
supernova mechanism.\,\cite{Retard,Leff} It has also been useful in 
deriving the Rayleigh-Plesset equation and various generalizations 
for the gas bubble dynamics in a liquid.\,\cite{RP,KTR} 
Furthermore, the global features of high-energy hadronic and nuclear collisions are described well by the fireball model, see Ref.\,\cite{fball} 
for a review. We aim at a dynamical scheme which improves the simplest fireball model in the direction of a realistic hydrodynamical description,
incorporating the QCD phase transition and important dissipative 
processes. 

In the present work, we reformulate {\it S}pecial and {\it G}eneral {\it R}elativistic (SR and GR, respectively) hydrodynamics as a variational
principle. We derive a SR generalization of the
Rayleigh-Plesset equation in this approach. Furthermore, 
we obtain the GR effective Lagrangian for spherically symmetric systems.

\section{The Variational Approach}

\subsection{Variables and Constraints}

While the variational formulation of nonrelativistic hydrodynamics has
been established for quite a while,   
we consider SR and GR equations of motion here. The most natural variable 
describing the flow is the velocity field, $\vec{v}(\vec{r},t)$. 
Introducing the four-vector $u^{\mu }(x)$, we define instead the 
variational variables,  
\begin{equation} \label{u}
u^{0}\equiv\gamma \;\;,\;\;\; \vec{u}\equiv\gamma\vec{v}
\;\;, \end{equation}  
i.e. functions of $\vec{r}$ and $t$, subject to the constraint, 
\begin{equation} \label{uu=1}
u_{\mu}u^{\mu}=1
\;\;. \end{equation}
The flow of matter provokes local changes in the occupied volume. 
Considering the case of a conserved chargelike quantity, say the 
baryon number, let
its local density in the comoving frame be $n$.
Then, 
\begin{equation} \label{conti}
\partial _{\mu}\left( nu^{\mu}\right)=0
\;\;. \end{equation}
The energy of the matter is given by: 
\begin{equation} \label{energy}
E=\varepsilon V \;\;,\;\;\; V=\frac{1}{n}
\;\;, \end{equation}
where $\varepsilon$ denotes the energy density and $V$ the {\it local}  specific 
volume. Assuming local
equilibrium, we have the thermodynamical relations, 
\begin{equation} \label{p}
\left( \frac{\partial E}{\partial V}\right) _{S}=-p 
\;\;,\;\;\;
\left( \frac{\partial \varepsilon }{\partial n}\right) _{S}=
\frac{\varepsilon +p}{n}
\;\;, \end{equation}
where $S$ is the entropy and $p$ the local pressure. 

\subsection{The SR Action and Hydrodynamic Equations of Motion}

Let us first
consider a single particle of rest mass $m_{0}$ with the action,  
\begin{equation} \label{actionP}
I_{P}=-m_0\int ds=-\int dt\gamma ^{-1}m_{0}=-\int dt\;\varepsilon dV
\;\;, \end{equation}
where $\gamma ^{-1}\equiv (1-v^2)^{1/2}$; here we assumed an infinitesimal rest frame volume $dV_0$ of the particle, such that 
$m_0\equiv\varepsilon dV_0=\varepsilon\gamma dV$ in the 
laboratory frame, where the particle has velocity $v$.    

This consideration leads to the action for a fluid, 
$I_M=-\int d^{4}x\,\;\varepsilon$, which is considered here  
as an aggregation of infinitesimal volume elements.

Varying the action with respect to the four-velocity field, 
$u^{\mu}\rightarrow u^{\mu}+\delta u^{\mu}$, and the density distribution, 
$n\rightarrow n+\delta n$, we have to respect the constraints, 
Eqs.\,(\ref{uu=1}),\,(\ref{conti}). Introducing Lagrange multipliers, 
we obtain the fluid action, 
\begin{equation} \label{L}
I_{M}=\int d^{4}x\left\{ -\varepsilon (n)+\xi (x)\partial _{\mu}
[nu^{\mu}]+\frac{1}{2}\zeta (x)[u^{\mu}u_{\mu}-1]\right\}
\;\;, \end{equation}
and the variational principle is stated as: 
$\delta I_{M}=0$, for arbitrary variations of 
$u^{\mu},\,n,\,\xi ,\,\zeta$. 

In practical applications of the variational principle it will be 
convenient to parametrize $u^{\mu },\,n $ such that the 
constraints are automatically satisfied and the Lagrange multipliers 
$\xi ,\,\zeta$ eliminated from the beginning. 

We remark that the following derivation of
the hydrodynamic equations of motion is equally valid in general
coordinate systems. Then, the partial derivative $\partial ^{\mu}$ 
should be replaced by the appropriate covariant derivative 
in the action and correspondingly in the following. Furthermore, the 
volume element $d^{4}x$ has to be replaced by the invariant volume element $\sqrt{-g}d^{4}x$.\,\cite{Weinberg}

Performing first a partial integration of the second term of the 
action, Eq.\,(\ref{L}), and then the four variations, 
we obtain the set of equations: 
\begin{eqnarray} 
\label{zetav} 
u^2&=&1
\;\;, \\
\label{xiv}
\partial_\mu (nu^\mu )&=&0
\;\;, \\
\label{nv}
\frac{\partial\varepsilon}{\partial n}+u^\mu \partial_\mu\xi &=&0
\;\;, \\ 
\label{uv}
n\partial_\mu \xi -\zeta u_\mu &=&0 
\;\;, \end{eqnarray} 
from which the Lagrange multipliers can be eliminated. 

In this calculation, it is useful to realize that 
$u^\mu \partial_\mu \equiv\partial_\tau$, i.e. the proper time derivative. 
Then, we calculate that 
$\partial_\mu \partial_\tau \xi =\partial_\tau \partial_\mu \xi$.
Furthermore, for adiabatic changes, cf. Eqs.\,(\ref{p}), we make use 
of the relation, 
\begin{equation} \label{adiab}
d\left( \frac{\varepsilon +p}{n}\right) =\frac{dp}{n}
\;\;. \end{equation}
With these ingredients, we finally obtain:  
\begin{equation} \label{ff}
(\varepsilon +p)u^\nu \partial_\nu u_\mu
+u_\mu u^\nu \partial_\nu p
=\partial_\mu p
\;\;, \end{equation}
which is the hydrodynamic four-vector equation of motion. 
 
Separating the space and time components of Eq.\,(\ref{ff}), 
recalling Eqs.\,(\ref{u}), we immediately obtain the results: 
\begin{eqnarray} 
\label{Econs}
[\partial_t+\vec{v}\cdot\nabla ]\gamma &=&
\frac{1}{\gamma (\varepsilon +p)}
(\partial_t p-\gamma ^2[\partial_t+\vec{v}\cdot\nabla ]p)
\;\;, \\
\label{Euler}
[\partial_t+\vec{v}\cdot\nabla ]\vec{v}&=&
-\frac{1}{\gamma ^2(\varepsilon +p)}(\nabla p+\vec{v}\partial_t p)
\;\;, \end{eqnarray} 
i.e. the relativistic energy and momentum flow equations, the latter 
being known as the SR generalization of the Euler 
equation.\,\cite{Weinberg}    

\subsection{Example: Relativistic Rayleigh-Plesset Equation for a
Spherical Bubble}

In all the applications of relativistic hydrodynamics to 
multiparticle production relatively little attention has been payed to the 
spherical geometry.\,\cite{Fermi,CF} The longitudinal 
expansion is usually considered predominant compared to the transverse one.\,\cite{Landau,Bjorken}. However, in collisions of very heavy nuclei
the transverse expansion might be equally important. Therefore, applying 
the variational principle, we derive an effective equation of motion 
for a spherically symmetric fluid bubble. We sketch the 
most simplified scenario, while more details can be found 
elsewhere.\,\cite{KTR}

Let $n_0(r,t)$ denote the baryon number density inside a bubble 
of radius $R(t)$ for a space fixed coordinate system (center-of-mass). In 
a comoving frame the density is $n=\gamma ^{-1}n_0$, 
with $\gamma ^{-1}=(1-\vec{v}^2)^{1/2}$. We make the {\it ansatz}, 
\begin{equation} \label{dens}
n_0(r,t)\equiv R^{-3}f(x;a(t))\;\;,\;\;\;x\equiv r/R
\;\;, \end{equation} 
where $f$ denotes the shape function depending parametrically on $a$  
(or several components $a_i$)   
and on the scaling variable $x$. Then, the continuity equation, 
$\partial_tn_0+r^{-2}\partial_r(r^2vn_0)=0$, can be solved for 
the radial velocity, 
\begin{equation} \label{velo}
v(x,t)=x\dot R-\frac{R\dot a}{x^2f}\int_0^xdx'\;x'^2\frac{\partial f}
{\partial a}
\;\;. \end{equation}  
Since the constraints incorporated in the action $I_M$, Eq.\,(\ref{L}), 
are automatically satisfied by now, the effective action becomes: 
\begin{equation} \label{Leff} 
I_{RP}\equiv\int dt\;L =-4\pi\int dt\;R^3\int_0^1dx\;x^2\varepsilon
\left( f(x;a)/(\gamma R^3)\right)
\;\;. \end{equation}
The corresponding  Euler-Lagrange equations present coupled 
ordinary differential equations for $R(t)$ and $a(t)$, which 
demonstrates the 
essential simplification due to the variational ansatz. 
 
In the limiting case of homologous motion, i.e. a time independent 
shape function $f(x)$ and consequently a linear velocity profile, 
$v(x,t)=x\dot R$, the resulting Rayleigh-Plesset 
type equation is: 
\begin{equation} \label{RP}
\frac{d}{dt}\left\{\dot RR^3\int_0^1dx\;x^4\gamma ^2(p+\varepsilon )\right\} 
=3R^2\int_0^1dx\;x^2p
\;\;, \end{equation} 
which fully incorporates the necessary SR corrections.  

Assuming an ultrarelativistic adiabatic equation of state for
an idealized quark-gluon plasma fluid, we obtain the relations, 
\begin{equation} \label{qgp} 
p=p_0[R_0/R(t)]^{3\Gamma}f^\Gamma (x)\gamma ^{-\Gamma}
\;\;,\;\;\; \varepsilon =\varepsilon_0p/p_0 
\;\;, \end{equation} 
with the adiabatic index $\Gamma =4/3$. The  
nonperturbative QCD vacuum pressure $B\approx (200\mbox{MeV})^4$ 
acting on such a plasma from the outside,\,\cite{review} 
is incorporated here by replacing $p\rightarrow p-B$ 
on the right hand side of Eq.\,(\ref{RP}).  
Then, it is straightforward to solve Eq.\,(\ref{RP}) together with 
Eq.\,(\ref{qgp}) numerically for suitable initial conditions, which  
illustrates our approach.\,\cite{KTR} 

Along these lines, also an expanding 
hadronic phase outside the plasma bubble and dissipative processes  
due to the phase boundary can be incorporated and are presently 
under study.      

\section{General Relativistic Hydrodynamics}
For most astrophysical or cosmological applications of the variational approach it will be necessary to incorporate GR gravity. 
According to our considerations in Sec.\,2.2, we have to combine 
the usual action for gravity, $I_G$, with a suitable fluid/matter action, 
$I_M$. The following total action suggests itself: 
\begin{equation} \label{GRaction}
I_G+I_M\equiv\int d^4x\sqrt{-g}\left\{ \frac{R}{16\pi G}
-\varepsilon (n)+\xi (x)\nabla_{\mu}
[nu^{\mu}]+\frac{1}{2}\zeta (x)[u^{\mu}u_{\mu}-1]\right\}
\;\;, \end{equation}
where the only difference in the matter part, cf. Eq.\,(\ref{L}), 
arises from co-/invariance requirements; we use 
standard notation.\,\cite{Weinberg} Here the variational principle 
is: $\delta S=0$, for all independent variations of the metric 
$g_{\mu\nu}$ and of $u^{\mu},\,n,\,\xi ,\,\zeta$, as before.

Performing the variations, using 
$\partial_\mu\ln\sqrt{-g} =\Gamma^\alpha_{\mu\alpha}$, we obtain  
Eqs.\,(\ref{zetav}), (\ref{nv}), and (\ref{uv}). Furthermore, 
varying $\xi$ and $g_{\mu\nu}$ leads to, respectively: 
\begin{eqnarray} 
\label{xiv1}
&\;&\nabla_\mu [nu^\mu ]\equiv (\partial_\mu +\Gamma^\alpha_{\mu\alpha})
[nu^\mu ]=0
\;\;, \\ 
\label{gv}
&\;&(8\pi G)^{-1}G^{\mu\nu}
-\frac{1}{2}\sqrt{-g}[(\varepsilon +p)u^\mu u^\nu -pg^{\mu\nu}]=0 
\;\;, \end{eqnarray} 
where $G^{\mu\nu}$ is the Einstein tensor, and we used 
$\partial_{g_{\mu\nu}}\ln\sqrt{-g}=-g^{\mu\nu}/2$. 

Identifying the 
energy-momentum tensor by 
$\delta I_M/\delta g_{\mu\nu}\equiv\sqrt{-g}T^{\mu\nu}/2$, we observe 
that Eq.\,(\ref{gv}) is nothing but Einstein's equation for an
ideal fluid, $G^{\mu\nu}=8\pi GT^{\mu\nu}$, with 
$T^{\mu\nu}=(\varepsilon +p)u^\mu u^\nu -pg^{\mu\nu}$. 
Then, the Bianchi identity, 
$\nabla_\mu G^{\mu\nu}=0$, implies the GR hydrodynamic equations: 
$\nabla_\mu T^{\mu\nu}=0$.\,\cite{Weinberg}
  
We conclude by briefly presenting the related effective 
Lagrangian for the most general
spherically symmetric system, where the metric is: 
\begin{equation} \label{metric}
ds^2=\mbox{e}^{2\phi}dt^2-\mbox{e}^{2\lambda}dr^2-R^2d\Omega^2
\;\;, \end{equation} 
with $\phi ,\,\lambda ,\,R$ unknown (variational) functions 
of $r,\, t$. Choosing the comoving frame with $u^\mu =(u^0,0,0,0)$,  
the normalization $u^2=1$ implying $u^0=\exp (-\phi )$, and with 
the (baryon) density $n$ being proportional to 
$R^{-2}\exp (-\lambda )$, the current conservation follows. Thus, 
the constraints being satisfied, the Lagrangian corresponding 
to Eq.\,(\ref{GRaction}) for spherical symmetry is obtained: 
\begin{equation} \label{Lagrange}
{\cal L}=
\mbox{e}^{\phi }\mbox{e}^{\lambda }R^{2}
[
-\varepsilon (n)+\frac{1}{16\pi GR^{2}}\{ R^{\prime}
[2R\phi^{\prime}+R^{\prime}]\mbox{e}^{-2\lambda}-\dot{R}
[2R\dot{\lambda}+\dot{R}]\mbox{e}^{-2\phi}+1\} 
]
\;, \end{equation}
where $\dot{f}\equiv\partial f/\partial t$ and 
$f^{\prime }\equiv\partial f/\partial r$; we omitted a total 
derivative which does not influence the 
related Euler-Lagrange equations for $\phi ,\,\lambda ,\,R$.

We have demonstrated that known cases like the Misner-Sharp equation 
relevant for stellar collapse or explosion, the 
Tolman-Oppenheimer-Volkov equation describing neutron star structure, 
and the Friedmann-Robertson-Walker cosmological model can be 
re-derived from this action. 

\section{Conclusions} 
All known results involving ideal fluids in SR and GR, respectively, can 
be obtained directly from the actions, Eqs.\,(\ref{L}) and (\ref{GRaction}), respectively, which we derived. 
Some earlier attempts can be found in the literature.\,\cite{actions}
However, we motivated in Sec.\,1 and stress here once more that the    
{\it application} of the variational principle together with physically 
motivated trial functions, as illustrated in Sec.\,2.3, may provide a 
powerful tool to study 
new and interesting fluids under extreme conditions, such as in ultrarelativistic nuclear 
collisions, or some more complex systems in GR.  
We will report the details of our calculations and such new 
applications elsewhere. 

\section*{Acknowledgments}
HTE thanks H. Fried for last-minute support, and 
him and B. M\"uller for the invitation to the  
stimulating QCD workshop. 
This work was supported in part by CNPq (Brasil),
PRONEX-41.96.0886.00,
FAPESP-95/4635-0, and by a grant from the U.S. Department of
Energy, DE-FG03-95ER40937.

\end{document}